\magnification=1200
\hoffset=.0cm
\voffset=.0cm
\baselineskip=.53cm plus .53mm minus .53mm

%
%
%
%
\def\ref#1{\lbrack#1\rbrack}
%
%
%
%
\input amssym.def
\input amssym.tex
%
%
\font\teneusm=eusm10                    
\font\seveneusm=eusm7                   
\font\fiveeusm=eusm5                 
%
%

%
%

%
%
\newfam\eusmfam
\textfont\eusmfam=\teneusm
\scriptfont\eusmfam=\seveneusm
\scriptscriptfont\eusmfam=\fiveeusm

\def\proclaim #1. #2\par{\medbreak{\bf #1.\enspace}{\it #2}\par\medbreak}
%
%
%
%
%

\def\hst1{\hskip 1pt}
%
%
%
%

\def\CN{{\cal N}}

\hbox to 16.5 truecm{March 2000   \hfil DFUB 00--05}
\hbox to 16.5 truecm{Version 1  \hfil hep-th/0003230}
\vskip2cm
\centerline{\bf 3-point functions of universal scalars in maximal  
SCFTs at large $N$} 
\vskip1cm
\centerline{by}
\vskip.5cm
\centerline{Fiorenzo Bastianelli and  Roberto Zucchini}
\centerline{\it Dipartimento di Fisica, 
Universit\`a degli Studi di Bologna}
\centerline{\it V. Irnerio 46, I-40126 Bologna, Italy}
\centerline{and}
\centerline{\it INFN, Sezione di Bologna}
\vskip1cm
\centerline{\bf Abstract} 

We compute all 3-point functions of the ``universal'' scalar 
operators contained in the interacting, maximally supersymmetric 
CFTs at large $N$ by using the AdS/CFT correspondence.
These SCFTs are related to the low energy description of 
M5, M2 and D3 branes, and the common set of universal scalars
corresponds through the AdS/CFT relation to the fluctuations of 
the metric and the magnetic potential along the internal manifold.
For the interacting $(0,2)$ SCFT$_6$ at large $N$,
which is related to M5 branes, this set of scalars is 
complete, while additional non-universal scalar operators 
are present in the $d=4$, $\CN=4$ super Yang--Mills theory 
and in the $\CN=8$ SCFT$_3$,
related to D3 and M2 branes, respectively.

\par\noindent
\vfill\eject
\par\vskip.6cm
\item{\bf 0.} {\bf Introduction}
\vskip.4cm
\par

The low energy dynamics of non-dilatonic superstring/M-theory branes
identify an interesting class of interacting CFT possessing 
maximal supersymmetry [1].
These SCFT should presumably be given a description in terms 
of the collective coordinates of the branes. 
This is well known for a system of $N$ coinciding D3 branes, whose  
collective degrees of freedom span the $d=4$, ${\cal N}=4$ $U(N)$ 
super vector multiplet and whose infrared dynamics is described
precisely by the corresponding super Yang--Mills theory [2].
A similar explicit description for a set of $N$ 
coinciding M5 or M2 branes is unknown.
Indeed, it is only known that the collective coordinates of a single M5 brane
form a $d=6$, ${\cal N}=(0,2)$ free tensor multiplet 
(a 2-form with selfdual field strength, 5 scalars and 2 Weyl fermions) [3,4] 
and that those of a single M2 brane form a $d=3$, ${\cal N}=8$ free
scalar multiplet (8 bosons and 8 Majorana fermions) [5].
Interacting SCFTs describing the collective coordinates
of $N$ coinciding M5 or M2 branes remain instead quite mysterious.  However,
superstring/M-theory predicts the existence of these  
models [6--8] and, in fact, suggests a full ADE classification.
A concrete handle on the problem is provided by the AdS/CFT duality
conjecture, which relates these SCFT
in $d=3,4,6$ to superstring/M-theory on AdS$_{4,5,7}\times$S$_{7,5,4}$ 
[9--11]. Specifically, 
at large $N$, one can approximate the superstring/M-theory by the 
corresponding classical supergravity. Then, the latter may be used to obtain 
informations on the strong coupling limit of the related 
interacting SCFTs and, in the case of SCFT$_{3,6}$, also to get 
important clues about their mysterious lagrangian formulation.
The AdS/CFT duality conjecture has been tested 
extensively in the literature (see ref. [12] and references therein)
and used to compute 3- and 4-point functions of some chiral operators
(a partial list consists of refs. [13--24]).

In this paper, we continue our analysis of the AdS/CFT correspondence 
presented in [22--24], and address the problem of computing
the 3 point functions for a set of scalar operators
which are present in all of the above mentioned theories
at large $N$. In fact, there are three families of such scalars,
to be denoted by ${\cal O}^s$, ${\cal O}^f$, ${\cal O}^t$, 
which are in correspondence with the metric and the magnetic potential 
fluctuations along the internal manifold 
(the metric contributing with two Kaluza-Klein families 
due to its splitting into a trace and a traceless part).
They form a kind of ``universal'' scalar sector 
common to all the non-dilatonic branes, which is reminiscent of the NS-NS
universal sector present in the spectrum of the various closed superstrings.
At large $N$, all but a finite number of the supersymmetric
short multiplets contain precisely 
one scalar ${\cal O}^s$, ${\cal O}^f$, ${\cal O}^t$, 
with ${\cal O}^s$ being the chiral primary.
Their conformal dimensions are given by
$$
\Delta^s=\Delta, \quad\quad\Delta^f=\Delta+2, 
\quad\quad\Delta^t=\Delta+4,
\eqno(0.1)
$$
where
$$
\Delta={n+1\over D-n-3}k,
\eqno(0.2)
$$
$n$ is the dimension of the brane,
$D$ is the space time dimension and $k\geq 4$ is an integer
characterizing the multiplet. The exceptional multiplets,
corresponding to the values $k=2,3$, do not contain the operator 
${\cal O}^t$.
For the M5 brane case, this set of scalars is complete at large $N$,
as all other operators have a non trivial tensorial character.
On the other hand, for the M2 and D3 branes,
additional non-universal scalars are present in the spectrum
(see, e.g. the tables in [25, 26]).

The simultaneous treatment of the universal sector is made possible
by the use of the general gravitational model 
introduced in [23], which was shown later in [24] to correctly reproduce
the universal scalar self couplings also in the case of the dyonic 
D3 brane (where the full lagrangian of type IIB supergravity contains 
a 4-form with selfdual field strength).

Thus, in the following, we briefly review our general gravitational 
model and present the result for all of the 3-point couplings
of the universal scalars. As a check, 
we have carried out the computation in two independent ways,
by expanding the action at the cubic order in the scalar fields
and by studying the quadratic corrections to the scalar equations of 
motion. Of course, we obtained the same final result.
Then, the application of AdS/CFT duality 
allows us to obtain the announced universal scalar
3-point functions. We tried to cast the resulting expressions   
in a way which may suggest a group theoretic interpretation.
Finally, we present some technical details on tensor spherical harmonics
and their integrals in an appendix.

\par\vskip.6cm
\item{\bf 1.} {\bf Identification of the universal scalar sector}
\vskip.4cm
\par
According to the AdS/CFT duality principles [9--11], 
the low energy world volume CFT of $N$ coincident M5, M2, D3 
branes at large $N$ is described by $D=11$, $D=11$, $D=10$ type IIB 
classical supergravity compactified on 
${\rm AdS}_7\times{\rm S}_4$, ${\rm AdS}_4\times{\rm S}_7$,
${\rm AdS}_5\times{\rm S}_5$, respectively. 
The CFT scalar fields,
whose three point functions we want to compute, 
are related by duality in a precise way 
to certain fluctuations of the AdS bosonic supergravity fields around a 
maximally supersymmetric Freund--Rubin background [27].
In ref. [23, 24], it has been shown that the dynamics of these 
fluctuations up to third order
is governed by a gravitational action that has the same form 
for all the three types of
branes mentioned above and is, therefore, universal.
Let us describe and justify the model briefly.

Space time $M_D$ is ${\rm AdS}_{D-2-p}\times {\rm S}_{2+p}$.
The relevant fields are the metric $g$ and the $1+p$ form field
$A_{1+p}$ with field strength $F_{2+p}=dA_{1+p}$
\footnote{}{}
\footnote{${}^1$}{We denote form degree on $M_D$ by a subfix, e. g. 
$\omega_r$ is a $r$ form on $M_D$.}.
The Freund--Rubin background $\bar g$, $\bar A_{1+p}$ is such 
that $\bar g$ is factorized and $\bar F_{2+p}=\bar F_{(0,2+p)}$. 
The relevant bosonic fluctuations of 
$g$ and $A_{1+p}$ around the background are such that $g$ remains 
factorized and $F_{2+p}=\bar F_{(0,2+p)}+da_{(0,1+p)}$
\footnote{}{}
\footnote{${}^2$}{ 
We say that a metric $g$ on $M_D$ is factorized if $g$ has the block 
structure $g=g'\oplus g''$, where $\bar g'$, $\bar g''$ are metrics
on ${\rm AdS}_{D-2-p}$, ${\rm S}_{2+p}$, respectively.
 Similarly, we denote form degree on the factors ${\rm AdS}_{D-2-p}$, 
${\rm S}_{2+p}$ by a pair of suffixes, e. g. $\omega_{(r,s)}$ denotes a 
$r+s$ form on $M_D$ that is a $r$ form on ${\rm AdS}_{D-2-p}$ and a $s$ form 
on ${\rm S}_{2+p}$.}. The action is given by 
$$
I={1\over 4\kappa^2}\int_{{\rm AdS}_{D-2-p}\times{\rm S}_{2+p}}
\Big[R(g)*_g1-F_{2+p}\wedge *_gF_{2+p}\Big],
\eqno(1.1)
$$
$$
g=g'\oplus g'',\quad F_{2+p}=\bar F_{(0,2+p)}+da_{(0,1+p)}.
\eqno(1.2)
$$
The important cases are those for which 
$(D,p)=(11,2),~(11,5),~(10,3)$. 

For the M5 theory ($(D,p)=(11,2)$), the above action is obtained directly 
from that of the bosonic sector of $D=11$ supergravity by noting that,
for the fluctuation considered here, the Chern--Simons term vanishes
identically.

For the M2 theory ($(D,p)=(11,5)$), the situation is slightly more 
complicated. Using the standard 3--form formulation of $D=11$ 
supergravity is inconvenient as the relevant scalar fluctuation
contained in the fluctuation $a_{(3,0)}$ of $A_3$ comes about as the 
solution of the constraint
$$
d*_{\bar g'}a_{(3,0)}=0
\eqno(1.3)
$$ 
entailed by gauge fixing at quadratic level, which is difficult to 
implement in an off--shell fashion. This problem can be solved by means of a
standard dualization trick, whereby the 3--form $A_3$ is replaced by
a 6--form $A_6$ such that $F_7=*_gF_4$. Since the Chern--Simons term 
vanishes also in this case for the fluctuation considered here, 
the resulting action takes the simple form (1.1). However, in the dual 
formulation, the relevant scalar fluctuation is contained in the fluctuation 
$a_{(0,6)}$ of $A_6$ and can thus be treated in an off--shell fashion
in a way completely analogous to that of the M5 brane. 

For the D3 theory ($(D,p)=(10,3)$), the relevant fields are the metric 
$g$ and the IIB Ramond--Ramond 4 form field $A_4$ with selfdual field 
strength $F^{sd}_5=dA_4$
$$
F^{sd}_5=*_gF^{sd}_5.
\eqno(1.4)
$$
The selfdual Freund--Rubin background $\bar g$, $\bar A_4$ 
is such that $\bar g$ is factorized and $\bar F^{sd}_5
=2^{-{1\over 2}}(\bar F_{(0,5)}+*_{\bar g}\bar F_{(0,5)})$.
The relevant fluctuations of $g$ and $A_4$ around the background
are such that $g$ is factorized, as usual, and $F^{sd}_5
=\bar F^{sd}_5+da_4$, where $a_4=2^{-{1\over 2}}(a_{(4,0)}+a_{(0,4)})$. 
The selfduality equations (1.4) relate the fluctuations 
$a_{(4,0)}$, $a_{(0,4)}$ and allow in principle to express 
$a_{(4,0)}$ in terms of $a_{(0,4)}$. Upon doing this, the resulting 
field equations can be seen to be equivalent to those obtained from the 
action (1.1).

The standard ${\rm AdS}_{D-2-p} \times {\rm S}_{2+p}$
Freund Rubin solution $\bar g_{ij}$, $\bar A_{i_1\cdots i_{1+p}}$ 
of the field equations following from the action (1.1) is given by
$$
\eqalignno{ 
\bar R_{\kappa\lambda\mu\nu}
&=-a^2(\bar g_{\kappa\mu}\bar g_{\lambda\nu}
-\bar g_{\kappa\nu}\bar g_{\lambda\mu}),  
\quad\quad a^2={(1+p)\over{(D-2)(D-3-p)}}e^2,&(1.5a)\cr
\bar R_{\alpha\beta\gamma\delta}
&=\bar e^2(\bar g_{\alpha\gamma}\bar g_{\beta\delta}
-\bar g_{\alpha\delta}\bar g_{\beta\gamma}),
\quad\quad \bar e^2 ={(D-3-p)\over (D-2)(1+p)}e^2,&(1.5b)\cr
\bar F_{\alpha_1\cdots\alpha_{2+p}}
&=e\bar \epsilon_{\alpha_1\cdots\alpha_{2+p}}, 
\vphantom{(D-3-p)\over (D-2)(1+p)}  &(1.6)\cr}
$$
where $\bar \epsilon_{\alpha_1\cdots\alpha_{2+p}}$ denotes the standard
volume form on the unit sphere and $e$ is an arbitrary mass scale 
parameterizing the compactification
\footnote{}{}
\footnote{${}^3$}{In this paper, we adopt the following conventions. 
Latin lower case letters $i,j,k,l,\ldots$ denote $M_D$ indices.
Late Greek lower case letters $\kappa,\lambda,\mu,\nu\dots$
denote ${\rm AdS}_{D-2-p}$ indices.
Early Greek lower case letters $\alpha,\beta,\gamma,\delta,\ldots$
denote ${\rm S}_{2+p}$ indices.}. The other components of the Riemann tensor 
and the field strength vanish identically.

We expand the action in fluctuations around the
background $\bar g_{ij}$, $\bar A_{i_1\cdots i_{1+p}}$.
We parameterize the fluctuations $\delta g_{ij}$, $\delta A_{i_1\cdots i_{1+p}}$
of the fields $g_{ij}$, $A_{i_1\cdots i_{1+p}}$ around the background 
as in [28]
$$
\eqalignno{
\delta g_{\alpha\beta}&=f_{\alpha\beta}+\bar\nabla_\alpha n_\beta
+\bar\nabla_\beta n_\alpha+(\bar\nabla_\alpha\bar\nabla_\beta
-\hbox{$1\over 2+p$}\bar g_{\alpha\beta}\bar\nabla^\gamma\bar\nabla_\gamma)q
+\hbox{$1\over 2+p$}\bar g_{\alpha\beta}\pi,\vphantom{\Big[}&\cr
f^\gamma{}_\gamma&=0, \quad \bar\nabla^\gamma f_{\gamma\alpha}=0,\quad
\bar\nabla^\gamma n_\gamma=0,\vphantom{\Big[}&(1.7)\cr}
$$
$$
\eqalignno{
\delta A_{\alpha_1\cdots\alpha_{1+p}}&=
(1+p)\bar\nabla_{[\alpha_1}a_{\alpha_2\cdots\alpha_{1+p}]}
+\bar\epsilon_{\alpha_1\cdots\alpha_{1+p}}{}^\gamma\bar\nabla_\gamma b,
\vphantom{\Big[}&\cr
\bar\nabla^\gamma a_{\gamma\alpha_3\cdots\alpha_{1+p}}&=0.&(1.8)\cr}
$$
Fluctuations of the other components of $g_{ij}$ and $A_{i_1\cdots i_{1+p}}$ 
can be disregarded as they are independent up to third order
from the ones we are interested in and therefore do not contribute to 
the relevant scalar 3-point couplings. 

The gauge can be partially fixed
by eliminating those gauge invariances which do not correspond
to the usual reparameterization and form gauge invariances
from the ${\rm AdS}_{D-2-p}$ perspective. This can be done
by imposing
$$
\bar\nabla^\beta\big(\delta g_{\beta\alpha}
-\hbox{$1\over 2+p$}\bar g_{\beta\alpha}\delta g^\gamma{}_\gamma\big)=0,
\eqno(1.9)
$$
$$
\bar\nabla^\beta\delta A_{i_1\cdots i_p\beta}=0,
\eqno(1.10)
$$
as shown in [28]. Fixing the gauge entails a number of constraints which 
must be disposed of as explained in [24].

There are three universal families of ${\rm AdS}_{D-2-p}$ scalar fields 
contained in the fluctuations listed above, which we denote as
$f_I$, $s_I$, $t_I$.
The scalar fields $f_I$ are defined by expanding
$f_{\alpha\beta}$ (cfr. eq. (1.7)) with respect to an orthonormal 
basis $\{Y^{(2)}_I\}$ of symmetric traceless transversal 2--tensor 
spherical harmonics of ${\rm S}_{2+p}$ (cfr. appendix A1) 
$$
f_{\alpha\beta}=\sum_If_IY^{(2)}_{I\alpha\beta}.
\eqno(1.11)
$$
The scalar fields $s_I$, $t_I$ are given by linear functionals 
of $\pi$, $b$ (cfr. eq. (1.7), (1.8)) 
non local in ${\rm S}_{2+p}$ defined as follows. 
One expands the scalar fields $\pi$, $b$ with respect to an orthonormal 
basis $\{Y^{(0)}_I\}$ of scalar spherical harmonics of ${\rm S}_{2+p}$ 
(cfr. appendix A1) 
$$
\pi=\sum_I\pi_IY^{(0)}_I,\quad b=\sum_I b_IY^{(0)}_I
\eqno(1.12a),(1.12b)
$$
and identifies the scalar mass eigenstates as given by [23, 24]
$$
\eqalignno{
s_I&={1\over 2k+1+p}\bigg({1\over 2(2+p)(D-3-p)}\pi_I
+{(-1)^p(k+1+p)\over (1+p)(D-2)}eb_I\bigg),&(1.13a)\cr
t_I&={1\over 2k+1+p}\bigg({1\over 2(2+p)(D-3-p)}\pi_I
-{(-1)^p k\over (1+p)(D-2)}eb_I\bigg).&(1.13b)\cr}
$$

It should be kept in mind that the range of the quantum numbers $I$ of $f_I$
differs from that of the quantum numbers $I$ of $s_I$, $t_I$, as these 
ranges parameterize orthogonal bases of spherical harmonics of different
tensorial rank. However, we shall use the same notation for these 
different quantum numbers for simplicity, as no confusion is possible.

\par\vskip.6cm
\item{\bf 2.} {\bf The cubic action of the universal scalar sector}
\vskip.4cm

\par
From the action (1.1), one can extract the couplings of the scalars
$f_I$, $s_I$ and $t_I$ defined in the previous section. 
After performing some field redefinitions, one finds that
their action to cubic order is given by
$$
I^{fst}_{[\leq 3]}={1\over 4\kappa^2}
\int_{{\rm AdS}_{d+1}} \hskip -.8cm d^{d+1}y (-\bar g_{d+1})^{1\over 2}
\bigg [\hbox{$1\over 2$}
\sum_i A_i \psi_i (\square -m_i{}^2)\psi_i 
+\hbox{$1\over 3!$} \sum_{ijk} G_{ijk}\psi_i\psi_j\psi_k \bigg ],
\eqno(2.1)
$$
where $d=D-3-p$, $\square$ denotes the d'Alembertian on AdS$_{d+1}$
and the index $i=\{I,a\}$ contains a flavor index $a$ for the 
$(f,s,t)$ types of fields, i.e. $\psi_i=\psi^a_I=(f_I,s_I,t_I)$.
The various constants appearing in the actions 
are given by the following expressions.
$$
\eqalignno{
A^f_I&=\hbox{$1\over 2$} z_I \bar e^{-2-p},\vphantom{1\over 2}
&(2.2a)\cr
A^s_I&={2\nu k(k-1)(2k+1+p)\over k+\gamma_s} z_I \bar e^{-2-p},
&(2.2b)\cr
A^t_I&={2\nu (k+1+p)(k+2+p)(2k+1+p)\over k+\gamma_t} z_I \bar e^{-2-p};
&(2.2c)\cr}
$$
$$
\eqalignno{
m^f_I{}^2&=k(k+1+p)\bar e^2, &(2.3a)\cr
m^s_I{}^2&=k(k-1-p)\bar e^2, &(2.3b)\cr
m^t_I{}^2&=(k+1+p)(k+2+2p)\bar e^2; &(2.3c)\cr}
$$
$$
\eqalignno{
G^{fff}_{I_1I_2I_3} &=
\big(\alpha +\hbox{$1\over 2$}(1+p)\big)a_{I_1 I_2 I_3}
\langle{\cal T}_{I_1}{\cal T}_{I_2}{\cal T}_{I_3}\rangle
\bar e^{-p}, 
&(2.4a)\cr
G^{ffs}_{I_1I_2I_3} 
&=4(D - 2)\,
{\alpha_1\alpha_2\big(\alpha_3 + {1\over 2}(1+p)\big)
\big(\alpha + {1\over 2}(1+p)\big)
\over k_3+\gamma_s} a_{I_1 I_2 I_3}
\langle{\cal T}_{I_1}{\cal T}_{I_2}{\cal C}_{I_3}\rangle
\bar e^{-p},&(2.4b)\cr
G^{fft}_{I_1I_2I_3} 
&=4(D - 2)\,
{\big(\alpha_1 + {1\over 2}(1+p)\big)\big(\alpha_2 + {1\over 2}(1+p)\big)
\alpha_3 (\alpha + 1 + p)
\over k_3+\gamma_t} 
a_{I_1 I_2 I_3}
\langle{\cal T}_{I_1}{\cal T}_{I_2}{\cal C}_{I_3}\rangle
\bar e^{-p},&\cr
&&(2.4c)\cr
G^{fss}_{I_1I_2I_3}
&=8\nu\,{\alpha_1 \big(\alpha_1 - {1\over 2}(1+p)\big)
\alpha \big(\alpha + {1\over 2}(1+p)\big)
\over(k_2+\gamma_s)(k_3+\gamma_s)} &\cr
&~~\times\bigg\{
(\alpha_1-1) \Big(\alpha + {1+p\over D-3-p}\Big)
+{\theta\over \nu} F^{fss}
\bigg\}
a_{I_1 I_2 I_3}\langle{\cal T}_{I_1}{\cal C}_{I_2}{\cal C}_{I_3}\rangle
\bar e^{-p},&(2.4d)\cr
G^{ftt}_{I_1I_2I_3}
&=8\nu\, {\alpha_1 (\alpha_1 + 1 + p)
\big(\alpha + {1\over 2}(1+p)\big)
\big(\alpha + {3\over 2}(1+p)\big)
\over(k_2+\gamma_t)(k_3+\gamma_t)} &\cr
&~~\times\bigg\{
\Big(\alpha_1 + {(1+p)(D-4-p)\over D-3-p}\Big)(\alpha + p + 2) 
+{\theta\over \nu} F^{ftt} 
\bigg\}
a_{I_1 I_2 I_3}\langle{\cal T}_{I_1}{\cal C}_{I_2}{\cal C}_{I_3}\rangle
\bar e^{-p},&\cr
&&(2.4e)\cr
G^{fst}_{I_1I_2I_3}
&=8\nu\, {\alpha_1 (\alpha_2 + 1 + p)\big(\alpha_3 - {1\over 2}(1+p)\big)
\big(\alpha + {1\over 2}(1+p)\big)
\over(k_2+\gamma_s)(k_3+\gamma_t)} &\cr
&~~\times\bigg\{
\Big(\alpha_2 + {(1+p)(D-4-p)\over D-3-p}\Big)(\alpha_3 -1)
+{\theta\over \nu}F^{fst} 
\bigg\}
a_{I_1 I_2 I_3}\langle{\cal T}_{I_1}{\cal C}_{I_2}{\cal C}_{I_3}\rangle
\bar e^{-p},&\cr
&&(2.4f)\cr
G^{sss}_{I_1I_2I_3}
&=32(D-2)\nu\,
{\alpha_1\alpha_2\alpha_3\big(\alpha - {1\over 2}(1+p)\big) 
\big(\alpha + {1\over 2}(1+p)\big)\over 
(k_1+\gamma_s)(k_2+\gamma_s)(k_3+\gamma_s)} &\cr
&~~\times
\bigg\{
(\alpha-1) \Big(\alpha + {1+p\over D-3-p}\Big) 
\Big(\alpha + {(1+p)(-D+4+2p)\over 2(D-2)}\Big) 
+{\theta\over \nu} F^{sss} \bigg\}
&\cr
&~~\times
a_{I_1 I_2 I_3}\langle{\cal C}_{I_1}{\cal C}_{I_2}{\cal C}_{I_3}\rangle
\bar e^{-p},
&(2.4g)\cr
G^{sst}_{I_1I_2I_3}
&=32(D-2)\nu\,
\,{\big(\alpha_1 +{1\over 2}(1+p)\big)\big(\alpha_2 +{1\over 2}(1+p)\big)
\alpha_3 (\alpha_3-1-p)\big(\alpha+{1\over 2}(1+p)\big)
\over (k_1+\gamma_s)(k_2+\gamma_s)(k_3+\gamma_t)}&\cr
&~~\times\bigg\{ (\alpha_3-1)
\Big(\alpha_3 + {(1+p)(-D+3+p)\over D-2}\Big)
\Big(\alpha_3 + {(1+p)(-D+4+p)\over D-3-p}\Big)
- {\theta\over \nu} F^{sst} \bigg\} &\cr
&~~\times
a_{I_1 I_2 I_3}\langle{\cal C}_{I_1}{\cal C}_{I_2}{\cal C}_{I_3}\rangle
\bar e^{-p},
&(2.4h)\cr
G^{tts}_{I_1I_2I_3}
&=32(D-2)\nu\, 
\,{\alpha_1 \alpha_2 \big(\alpha_3 +{1\over 2}(1+p)\big)  
\big(\alpha_3 + {3\over2}(1+p)\big)(\alpha+1+p)
\over (k_1+\gamma_t)(k_2+\gamma_t)(k_3+\gamma_s)}&\cr
&~~\times
\bigg\{ (\alpha_3+2+p)
\Big(\alpha_3 + {(1+p)(3D-8-2p)\over 2 (D-2)}\Big)
\Big(\alpha_3 + {(1+p)(D-4-p)\over D-3-p}\Big)
- {\theta\over \nu}F^{tts} \bigg\}
&\cr
&~~\times
a_{I_1 I_2 I_3}\langle{\cal C}_{I_1}{\cal C}_{I_2}{\cal C}_{I_3}\rangle
\bar e^{-p},
&(2.4i)\cr
G^{ttt}_{I_1I_2I_3}
&=32(D-2)\nu\,
{\big(\alpha_1 + {1\over 2}(1+p)\big)
\big(\alpha_2 + {1\over 2}(1+p)\big)\big(\alpha_3+ {1\over 2}(1+p)\big) 
(\alpha + 1 + p) (\alpha +2 + 2p)\over (k_1+\gamma_t)(k_2+\gamma_t)(k_3+\gamma_t)} &\cr
&~~\times\bigg\{
(\alpha + 2 +p) \Big(\alpha + {(1+p)(2D-8-2p)\over D-3-p}\Big)
\Big(\alpha + {(1+p)(4D - 9 - p)\over 2(D-2)}\Big)
+{\theta\over \nu} F^{ttt} \bigg\}
&\cr
&~~\times
a_{I_1 I_2 I_3}\langle{\cal C}_{I_1}{\cal C}_{I_2}{\cal C}_{I_3}\rangle
\bar e^{-p},
&(2.4j)\cr}
$$
where we have defined for notational convenience
$$
\eqalignno{
\nu&=(D-2)(1+p)(D-3-p),\vphantom{2\over 3}&(2.5a)\cr 
\gamma_s&={1+p\over D-3-p},\quad\quad 
\gamma_t={(1+p)(D-4-p)\over D-3-p}, \vphantom{2\over 3} &(2.5b)\cr
\theta&=(1+p)(D-3-p)-2(D-2),\vphantom{2\over 3} &(2.5c)\cr}
$$
$\bar e^2$ is defined in eq. (1.5b), 
while $\alpha_1$, $\alpha_2$, $\alpha_3$, $\alpha$, 
$z_I$, $a_{I_1 I_2 I_3}$ and the contractions 
$\langle{\cal T}{\cal T}{\cal T}\rangle$, 
$\langle{\cal T}{\cal T}{\cal C}\rangle$,
$\langle{\cal T}{\cal C}{\cal C}\rangle$,
$\langle{\cal C}{\cal C}{\cal C}\rangle$ are defined in 
appendix A1.
Note that the auxiliary functions $F^{fss}, F^{ftt},...$  
appearing in the couplings are always multiplied by
$\theta$ and do not contribute to the relevant cases
of AdS$_{5,7,4}\times$S$_{5,4,7}$ where $\theta$ vanishes.
The explicit expressions for such functions, 
which may be needed in future applications, are
listed in appendix A2.

Writing above
$$
A_I=\bar A_I z_I \bar e^{-2-p},
\eqno(2.6)
$$
$$
m_I{}^2=\bar m_I{}^2 \bar e^2,
\eqno(2.7)
$$
$$
G_{I_1I_2I_3}=\bar G_{I_1I_2I_3}a_{I_1 I_2 I_3}
\langle{\cal Y}_{I_1}{\cal Y}_{I_2}{\cal Y}_{I_3}\rangle
\bar e^{-p},
\eqno(2.8)
$$
where the $\langle{\cal Y}_{I_1}{\cal Y}_{I_2}{\cal Y}_{I_3}\rangle$
denotes the appropriate tensor/scalar harmonic contractions, we get
the following expressions.

\vfill\eject

$\underline{{\rm AdS}_7\times{\rm S}_4}$

$$
\eqalignno{
\bar A^f_I&={1\over 2}, &(2.9a)\cr
\bar A^s_I&={2^2 3^4 k(k-1)(2k+3)\over k+{1\over 2}},
&(2.9b)\cr
\bar A^t_I&={2^2 3^4 (k+3)(k+4)(2k+3)\over k+{5\over 2}};
&(2.9c)\cr}
$$
$$
\eqalignno{
\bar m^f_I{}^2&=k(k+3),&(2.10a)\cr
\bar m^s_I{}^2&=k(k-3),&(2.10b)\cr
\bar m^t_I{}^2&=(k+3)(k+6);&(2.10c)\cr}
$$
$$\eqalignno{
\bar G^{fff}_{I_1I_2I_3}&= \alpha+ \hbox{$3\over2$},
&(2.11a)\cr
\bar G^{ffs}_{I_1I_2I_3}&= {2^2 3^2
\alpha_1\alpha_2(\alpha_3+{3\over 2})(\alpha+{3\over 2})
\over k_3+{1\over 2}},
&(2.11b)\cr 
\bar G^{fft}_{I_1I_2I_3}&= {2^2 3^2
(\alpha_1+{3\over 2})(\alpha_2+{3\over 2})\alpha_3(\alpha+3)
\over k_3+{5\over 2}}, 
&(2.11c)\cr
\bar G^{fss}_{I_1I_2I_3}&={2^4 3^4
(\alpha_1-{3\over 2})(\alpha_1-1)\alpha_1\alpha(\alpha+{1\over 2})
(\alpha+{3\over 2})\over (k_2+{1\over 2})(k_3+{1\over 2})},
&(2.11d)\cr
\bar G^{ftt}_{I_1I_2I_3}&={2^4 3^4
\alpha_1(\alpha_1+{5\over 2})(\alpha_1+3)(\alpha+{3\over 2})(\alpha+4)
(\alpha+{9\over 2})\over (k_2+{5\over 2})(k_3+{5\over 2})},
&(2.11e)\cr
\bar G^{fst}_{I_1I_2I_3}&={2^4 3^4
\alpha_1(\alpha_2+{5\over 2})(\alpha_2+3)(\alpha_3-1)
(\alpha_3-{3\over 2})
(\alpha+{3\over 2})\over (k_2+{1\over 2})(k_3+{5\over 2})},
&(2.11f)\cr
\bar G^{sss}_{I_1I_2I_3}&={2^6 3^6
\alpha_1\alpha_2\alpha_3
(\alpha-1)(\alpha^2-{1\over4})(\alpha^2-{9\over 4})
\over (k_1+{1\over 2})(k_2+{1\over 2})(k_3+{1\over 2})},
&(2.11g)\cr
\bar G^{sst}_{I_1I_2I_3}&={2^6 3^6
(\alpha_1+{3\over 2})(\alpha_2+{3\over 2})\alpha_3 
(\alpha_3-1)(\alpha_3-2)(\alpha_3-{5\over2})(\alpha_3-3)(\alpha+{3\over2})
\over (k_1+{1\over 2})(k_2+{1\over 2})(k_3+{5\over 2})},
&(2.11h)\cr
\bar G^{tts}_{I_1I_2I_3}&={2^6 3^6
\alpha_1\alpha_2(\alpha_3+{3\over 2})(\alpha_3+{5\over 2})
(\alpha_3+{7\over 2})(\alpha_3+4)(\alpha_3+{9\over 2})(\alpha+3)
\over (k_1+{5\over 2})(k_2+{5\over 2})(k_3+{1\over 2})},
&(2.11i)\cr
\bar G^{ttt}_{I_1I_2I_3}&= {2^6 3^6
(\alpha_1+{3\over 2})(\alpha_2+{3\over 2})(\alpha_3+{3\over 2}) 
(\alpha+3)(\alpha+4)(\alpha+5)(\alpha+{11\over 2})(\alpha+6)
\over (k_1+{5\over 2})(k_2+{5\over 2})(k_3+{5\over 2})}.&\cr
&&(2.11j)\cr}
$$

\vfill \eject 

$\underline{{\rm AdS}_4\times{\rm S}_7}$

$$
\eqalignno{
\bar A^f_I&={1\over 2}, &(2.12a)\cr
\bar A^s_I&={2^2 3^4 k(k-1)(2k+6)\over k+2},
&(2.12b)\cr
\bar A^t_I&={2^2 3^4 (k+6)(k+7)(2k+6)\over k+4};
&(2.12c)\cr}
$$
$$
\eqalignno{
\bar m^f_I{}^2&=k(k+6),&(2.13a)\cr
\bar m^s_I{}^2&=k(k-6),&(2.13b)\cr
\bar m^t_I{}^2&=(k+6)(k+12);&(2.13c)\cr}
$$
$$\eqalignno{
\bar G^{fff}_{I_1I_2I_3}&=\alpha+3,
&(2.14a)\cr
\bar G^{ffs}_{I_1I_2I_3}&=
{2^2 3^2 \alpha_1\alpha_2(\alpha_3+3)(\alpha+3)\over k_3+2},
&(2.14b)\cr
\bar G^{fft}_{I_1I_2I_3}&=
{2^2 3^2 (\alpha_1+3)(\alpha_2+3)\alpha_3(\alpha+6)\over k_3+4},
&(2.14c)\cr
\bar G^{fss}_{I_1I_2I_3}&=
{2^4 3^4 (\alpha_1-3)(\alpha_1-1)\alpha_1\alpha(\alpha+2)
(\alpha+3)\over (k_2+2)(k_3+2)},
&(2.14d)\cr
\bar G^{ftt}_{I_1I_2I_3}&=
{2^4 3^4 \alpha_1(\alpha_1+4)(\alpha_1+6)(\alpha+3)(\alpha+7)
(\alpha+9)\over (k_2+4)(k_3+4)},
&(2.14e)\cr
\bar G^{fst}_{I_1I_2I_3}&=
{2^4 3^4 \alpha_1(\alpha_2+4)(\alpha_2+6)(\alpha_3-3)
(\alpha_3-1)(\alpha+3)\over (k_2+2)(k_3+4)},
&(2.14f)\cr
\bar G^{sss}_{I_1I_2I_3}&=
{2^6 3^6 \alpha_1\alpha_2\alpha_3(\alpha+2)(\alpha^2-1)(\alpha^2-9)
\over (k_1+2)(k_2+2)(k_3+2)},
&(2.14g)\cr
\bar G^{sst}_{I_1I_2I_3}&=
{2^6 3^6 (\alpha_1+3)(\alpha_2+3)\alpha_3 
(\alpha_3-1)(\alpha_3-2)(\alpha_3-4)(\alpha_3-6)(\alpha+3)
\over (k_1+2)(k_2+2)(k_3+4)},
&(2.14h)\cr
\bar G^{tts}_{I_1I_2I_3}&=
{2^6 3^6 \alpha_1\alpha_2(\alpha_3+3)
(\alpha_3+4)(\alpha_3+5)(\alpha_3+7)(\alpha_3+9)(\alpha+6)
\over (k_1+4)(k_2+4)(k_3+2)},
&(2.14i)\cr
\bar G^{ttt}_{I_1I_2I_3}&=
{2^6 3^6 (\alpha_1+3)(\alpha_2+3)(\alpha_3+3) 
(\alpha+6)(\alpha+7)(\alpha+8)(\alpha+10)(\alpha+12)
\over (k_1+4)(k_2+4)(k_3+4)}.&\cr
&&(2.14j)\cr}
$$

\vfill \eject

$\underline{{\rm AdS}_5\times{\rm S}_5}$

$$
\eqalignno{
\bar A^f_I&={1\over 2}, &(2.15a)\cr
\bar A^s_I&={2^8 k(k-1)(2k+4)\over k+1},
&(2.15b)\cr
\bar A^t_I&={2^8 (k+4)(k+5)(2k+4)\over k+3};
&(2.15c)\cr}
$$
$$
\eqalignno{
\bar m^f_I{}^2&=k(k+4),&(2.16a)\cr
\bar m^s_I{}^2&=k(k-4),&(2.16b)\cr
\bar m^t_I{}^2&=(k+4)(k+8);&(2.16c)\cr}
$$
$$\eqalignno{
\bar G^{fff}_{I_1I_2I_3}&=\alpha+2,
&(2.17a)\cr
\bar G^{ffs}_{I_1I_2I_3}&=
{2^5\alpha_1\alpha_2(\alpha_3+2)(\alpha+2)\over k_3+1},
&(2.17b)\cr
\bar G^{fft}_{I_1I_2I_3}&=
{2^5(\alpha_1+2)(\alpha_2+2)\alpha_3(\alpha+4)\over k_3+3},
&(2.17c)\cr
\bar G^{fss}_{I_1I_2I_3}&=
{2^{10} (\alpha_1-2)(\alpha_1-1)\alpha_1\alpha(\alpha+1)
(\alpha+2)\over (k_2+1)(k_3+1)},
&(2.17d)\cr
\bar G^{ftt}_{I_1I_2I_3}&=
{2^{10} \alpha_1(\alpha_1+3)(\alpha_1+4)(\alpha+2)(\alpha+5)
(\alpha+6)\over (k_2+3)(k_3+3)},
&(2.17e)\cr
\bar G^{fst}_{I_1I_2I_3}&=
{2^{10} \alpha_1(\alpha_2+3)(\alpha_2+4)(\alpha_3-2)
(\alpha_3-1)(\alpha+2)\over (k_2+1)(k_3+3)},
&(2.17f)\cr
\bar G^{sss}_{I_1I_2I_3}&=
{2^{15} \alpha_1\alpha_2\alpha_3\alpha(\alpha^2-1)(\alpha^2-4)
\over (k_1+1)(k_2+1)(k_3+1)},
&(2.17g)\cr
\bar G^{sst}_{I_1I_2I_3}&=
{2^{15} (\alpha_1+2)(\alpha_2+2)\alpha_3 
(\alpha_3-1)(\alpha_3-2)(\alpha_3-3)(\alpha_3-4)(\alpha+2)
\over (k_1+1)(k_2+1)(k_3+3)},
&(2.17h)\cr
\bar G^{tts}_{I_1I_2I_3}&=
{2^{15} \alpha_1\alpha_2(\alpha_3+2)
(\alpha_3+3)(\alpha_3+4)(\alpha_3+5)(\alpha_3+6)(\alpha+4)
\over (k_1+3)(k_2+3)(k_3+1)},
&(2.17i)\cr
\bar G^{ttt}_{I_1I_2I_3}&=
{2^{15} (\alpha_1+2)(\alpha_2+2)(\alpha_3+2) 
(\alpha+4)(\alpha+5)(\alpha+6)(\alpha+7)(\alpha+8)
\over (k_1+3)(k_2+3)(k_3+3)}.&\cr
&&(2.17j)\cr}
$$

\vfill \eject


\par\vskip.6cm
\item{\bf 3.} {\bf Computation of the 3 point functions of the 
universal scalar sector}
\vskip.4cm

\par
We are now ready to compute two and three point functions
in the SCFTs using the ${\rm AdS}_{d+1}/{\rm CFT}_d$ correspondence.
The general formulas derived in [29,30] work with AdS radius set to 1.
Assume that the AdS scalar fields $\phi_i$ correspond to the CFT 
local field ${\cal O}_i$. The mass $m_i$ of $\psi_i$ and the 
conformal dimension $\Delta_i$ of ${\cal O}_i$ are related as
$$
\Delta_i=\hbox{$1\over2$}\Big[d+\Big(d^2+4m_i{}^2\Big)^{1\over 2}\Big].
\eqno(3.1)
$$
Then
$$
\langle {\cal O}_i(x) {\cal O}_j(y)\rangle = 
{A_i\over 4 \kappa^2} {2\over \pi^{d\over2}} 
{\Delta_i - {d\over 2} \over \Delta_i} 
{\Gamma(\Delta_i +1)\over \Gamma(\Delta_i -{d\over2})}
{(w_i)^2 \delta_{ij} \over |x-y|^{2\Delta_i}},
\eqno(3.2)
$$
where $ {A_i\over 4 \kappa^2} $ is the coefficient of the canonically 
normalized kinetic term of the bulk field $\psi_i$ as in eq. (2.1)
and 
$$
\langle {\cal O}_i(x) {\cal O}_j(y) {\cal O}_k(z) \rangle = 
{R_{ijk}\over |x-y|^{\Delta_i+\Delta_j-\Delta_k}
|y-z|^{\Delta_j+\Delta_k-\Delta_i}
|z-x|^{\Delta_k+\Delta_i-\Delta_j}},
\eqno(3.3)
$$
with
$$
\eqalignno{
R_{ijk}=& {G_{ijk}\over 4\kappa^2}{1\over 2\pi^d}
{\Gamma({1\over2}(\Delta_i+\Delta_j-\Delta_k))
\Gamma({1\over2}(\Delta_j+\Delta_k-\Delta_i))
\Gamma({1\over2}(\Delta_k+\Delta_i-\Delta_j))
\over 
\Gamma(\Delta_i -{d\over2})
\Gamma(\Delta_j -{d\over2})
\Gamma(\Delta_k -{d\over2})} 
&\cr 
&\Gamma(\hbox{${1\over2}$}({\Delta_i+\Delta_j+\Delta_k} - d))
w_iw_jw_k,&(3.4)\cr}
$$
where ${G_{ijk}\over 4\kappa^2}$ is the cubic coupling constant of 
$\psi_i$, $\psi_j$, $\psi_k$ as in eq. (2.1).
The factors $w_i$ parameterize unknown proportionality constants
which relate the fields $\psi_i$ to the sources of the operators
${\cal O}_i$, namely the generating functional
of correlators reads as 
$\langle e^{\int w_i \psi_i {\cal O}_i} \rangle_{_{\rm SCFT}}$. 
For the present purposes we follow ref. [13]
and fix them to normalize the two point functions as
$$
\langle {\cal O}_i(x) {\cal O}_j(y) \rangle = 
{\delta_{ij} \over |x-y|^{2\Delta_i}}.
\eqno(3.5)
$$
With this canonical normalization the three point functions are readily 
computed.

In our case, $d=D-3-p$.
Imposing that the AdS radius is 1 fixes the value of $\bar e$ to be
$$
\bar e={d\over 1+p}.
\eqno(3.6)
$$
We denote by ${\cal O}^f_I$, ${\cal O}^s_I$, ${\cal O}^t_I$ the ${\rm CFT}_d$
operators corresponding to the ${\rm AdS}_{d+1}$ scalars $f_I$, 
$s_I$, $t_I$ in 
the ${\rm AdS}_{d+1}/{\rm CFT}_d$  duality. Their dimensions are given by
$$
\eqalignno{
\Delta^f_I&={d(k+1+p)\over 1+p},&(3.7a)\cr
\Delta^s_I&={dk\over 1+p},&(3.7b)\cr
\Delta^t_I&={d(k+2+2p)\over 1+p}.&(3.7c)\cr}
$$

Separating out the group theory factors in the $R_{ijk}$ coefficients
by defining
$$
R_{ijk} = \bar R_{ijk}
\langle{\cal Y}_{I_1}{\cal Y}_{I_2}{\cal Y}_{I_3}\rangle
\eqno(3.8)
$$
as in eq. (2.8), we find the following expressions.

\vskip.2cm

$\underline{{\rm AdS}_7\times{\rm S}_4}$

\vskip.2cm

In this case one has ${1\over 4\kappa^2} = {2 N^3 \over \pi^5}$ and $\bar e=2$.
One has
$$
\eqalignno{
\Delta^f_I&=2k+6,&(3.9a)\cr
\Delta^s_I&=2k,&(3.9b)\cr
\Delta^t_I&=2k+12.&(3.9c)\cr}
$$
Set 
$$
\eqalignno{
\phi(k)&=4\bigg[{(2k+1)!\over (2k+2)!(2k+5)!}\bigg]^{{1\over 2}},&(3.10a)\cr
\sigma(k)&=\big[(2k-2)!\big]^{-{1\over 2}},&(3.10b)\cr
\tau(k)&=16\bigg[{(2k+1)!(2k+4)!(2k+7)!\over 
(2k+6)!(2k+8)!(2k+9)!(2k+11)!}\bigg]^{{1\over 2}}.
&(3.10c)\cr}
$$
Then,
$$
\eqalignno{
\bar R^{fff}_{I_1I_2I_3}&={2 \over 4(\pi N)^{3\over 2}}
\phi(k_1)\phi(k_2)\phi(k_3)
2^{2\alpha}{\Gamma(2\alpha+4)\Gamma(2\alpha+6)\over
\Gamma(2\alpha+3)\Gamma(2\alpha+5)}\Gamma(\alpha+3)\cr
&\times\vphantom{{1\over 4(\pi N)^{3\over 2}}}
{\Gamma(2\alpha_1+3)\Gamma(2\alpha_2+3)\Gamma(2\alpha_3+3)
\over \Gamma(2\alpha_1+2)\Gamma(2\alpha_2+2)\Gamma(2\alpha_3+2)}
\Gamma(\alpha_1+\hbox{$3\over 2$})\Gamma(\alpha_2+\hbox{$3\over 2$})
\Gamma(\alpha_3+\hbox{$3\over 2$})
&(3.11a)\cr
\bar R^{ffs}_{I_1I_2I_3}&={1\over 4(\pi N)^{3\over 2}}
\phi(k_1)\phi(k_2)\sigma(k_3)2^{2\alpha}\Gamma(\alpha+2)&\cr
&\times\vphantom{{1\over 4(\pi N)^{3\over 2}}}
{\Gamma(2\alpha_3+6)\over \Gamma(2\alpha_3+2)}
\Gamma(\alpha_1+\hbox{$1\over 2$})\Gamma(\alpha_2+\hbox{$1\over 2$})
\Gamma(\alpha_3+\hbox{$5\over 2$}),
&(3.11b)\cr
\bar R^{fft}_{I_1I_2I_3}&={1\over 4(\pi N)^{3\over 2}}
\phi(k_1)\phi(k_2)\tau(k_3)
2^{2\alpha}{\Gamma(2\alpha+9)\over\Gamma(2\alpha+5)}\Gamma(\alpha+4)&\cr
&\times\vphantom{{1\over 4(\pi N)^{3\over 2}}}
{\Gamma(2\alpha_1+6)\Gamma(2\alpha_2+6)
\over \Gamma(2\alpha_1+2)\Gamma(2\alpha_2+2)}
\Gamma(\alpha_1+\hbox{$5\over 2$})\Gamma(\alpha_2+\hbox{$5\over 2$})
\Gamma(\alpha_3+\hbox{$1\over 2$}),
&(3.11c)\cr
\bar R^{fss}_{I_1I_2I_3}&={1\over 4(\pi N)^{3\over 2}}
\phi(k_1)\sigma(k_2)\sigma(k_3)
2^{2\alpha}\Gamma(\alpha+1)&\cr
&\times\vphantom{{1\over 4(\pi N)^{3\over 2}}}
{\Gamma(2\alpha_2+3)\Gamma(2\alpha_3+3)
\over \Gamma(2\alpha_2+2)\Gamma(2\alpha_3+2)}
\Gamma(\alpha_1-\hbox{$1\over 2$})\Gamma(\alpha_2+\hbox{$3\over 2$})
\Gamma(\alpha_3+\hbox{$3\over 2$}),
&(3.11d)\cr
\bar R^{fst}_{I_1I_2I_3}&={1\over 4(\pi N)^{3\over 2}}
\phi(k_1)\sigma(k_2)\tau(k_3)
2^{2\alpha}{\Gamma(2\alpha+4)\Gamma(2\alpha+6)
\over \Gamma(2\alpha+3)\Gamma(2\alpha+5)}\Gamma(\alpha+3)&\cr
&\times\vphantom{{1\over 4(\pi N)^{3\over 2}}}
{\Gamma(2\alpha_1+1)\Gamma(2\alpha_1+3)
\over \Gamma(2\alpha_1)\Gamma(2\alpha_1+2)}
{\Gamma(2\alpha_2+3)\Gamma(2\alpha_2+7)\Gamma(2\alpha_2+9)
\over \Gamma(2\alpha_2+2)\Gamma(2\alpha_2+4)\Gamma(2\alpha_2+6)}
{\Gamma(2\alpha_3)\over\Gamma(2\alpha_3+1)}&\cr
&\times\vphantom{{1\over 4(\pi N)^{3\over 2}}}
\Gamma(\alpha_1+\hbox{$3\over 2$})\Gamma(\alpha_2+\hbox{$7\over 2$})
\Gamma(\alpha_3-\hbox{$1\over 2$}),
&(3.11e)\cr
\bar R^{ftt}_{I_1I_2I_3}&={1\over 4(\pi N)^{3\over 2}}
\phi(k_1)\tau(k_2)\tau(k_3)&\cr
&\times\vphantom{{1\over 4(\pi N)^{3\over 2}}}
2^{2\alpha}{\Gamma(2\alpha+4)\Gamma(2\alpha+6)
\Gamma(2\alpha+10)\Gamma(2\alpha+12)
\over\Gamma(2\alpha+3)\Gamma(2\alpha+5)\Gamma(2\alpha+7)\Gamma(2\alpha+9)}
\Gamma(\alpha+5)&\cr
&\times\vphantom{{1\over 4(\pi N)^{3\over 2}}}
{\Gamma(2\alpha_1+1)\Gamma(2\alpha_1+3)\Gamma(2\alpha_1+7)\Gamma(2\alpha_1+9)
\over
\Gamma(2\alpha_1)\Gamma(2\alpha_1+2)\Gamma(2\alpha_1+4)\Gamma(2\alpha_1+6)}
{\Gamma(2\alpha_2+3)\over\Gamma(2\alpha_2+2)}
{\Gamma(2\alpha_3+3)\over\Gamma(2\alpha_3+2)}&\cr
&\times\vphantom{{1\over 4(\pi N)^{3\over 2}}}
\Gamma(\alpha_1+\hbox{$7\over 2$})\Gamma(\alpha_2+\hbox{$3\over 2$})
\Gamma(\alpha_3+\hbox{$3\over 2$}),
&(3.11f)\cr
\bar R^{sss}_{I_1I_2I_3}&={1\over 4(\pi N)^{3\over 2}}
\sigma(k_1)\sigma(k_2)\sigma(k_3)
2^{2\alpha}\Gamma(\alpha)
\Gamma(\alpha_1+\hbox{$1\over 2$})
\Gamma(\alpha_2+\hbox{$1\over 2$})\Gamma(\alpha_3+ \hbox{$1\over 2$}),
&(3.11g)\cr
\bar R^{sst}_{I_1I_2I_3}&={1\over 4(\pi N)^{3\over 2}}
\sigma(k_1)\sigma(k_2)\tau(k_3)
2^{2\alpha}\Gamma(\alpha+2)&\cr
&\times
{\Gamma(2\alpha_1+6)\Gamma(2\alpha_2+6)
\over \Gamma(2\alpha_1+2)\Gamma(2\alpha_2+2)}
\Gamma(\alpha_1+\hbox{$5\over 2$})\Gamma(\alpha_2+\hbox{$5\over 2$})
\Gamma(\alpha_3-\hbox{$3\over 2$}),
&(3.11h)\cr
\bar R^{tts}_{I_1I_2I_3}&={1\over 4(\pi N)^{3\over 2}}
\tau(k_1)\tau(k_2)\sigma(k_3)
2^{2\alpha}{\Gamma(2\alpha+9)\over\Gamma(2\alpha+5)}\Gamma(\alpha+4)&\cr
&\times
{\Gamma(2\alpha_3+10)\Gamma(2\alpha_3+12)
\over \Gamma(2\alpha_3+2)\Gamma(2\alpha_3+8)}
\Gamma(\alpha_1+\hbox{$1\over 2$})\Gamma(\alpha_2+\hbox{$1\over 2$})
\Gamma(\alpha_3+\hbox{$9\over 2$}),
&(3.11i)\cr
\bar R^{ttt}_{I_1I_2I_3}&={1\over 4(\pi N)^{3\over 2}}
\tau(k_1)\tau(k_2)\tau(k_3)
2^{2\alpha}{\Gamma(2\alpha+13)\Gamma(2\alpha+15)
\over \Gamma(2\alpha+5)\Gamma(2\alpha+11)}
\Gamma(\alpha+6)&\cr
&\times
{\Gamma(2\alpha_1+6)\Gamma(2\alpha_2+6)\Gamma(2\alpha_3+6)
\over \Gamma(2\alpha_1+2)\Gamma(2\alpha_2+2)\Gamma(2\alpha_3+2)}
\Gamma(\alpha_1+\hbox{$5\over 2$})\Gamma(\alpha_2+\hbox{$5\over 2$})
\Gamma(\alpha_3+\hbox{$5\over 2$}).
&(3.11j)\cr}
$$
\vskip.2cm

$\underline{{\rm AdS}_4\times{\rm S}_7}$

\vskip.2cm

In this case one has ${1\over 4\kappa^2} = {N^{3\over 2} 
\over 2^{19\over 2} \pi^5}$ and $\bar e={1\over 2}$.
One has
$$
\eqalignno{
\Delta^f_I&=\hbox{$1\over 2$}k+3,&(3.12a)\cr
\Delta^s_I&=\hbox{$1\over 2$}k,&(3.12b)\cr
\Delta^t_I&=\hbox{$1\over 2$}k+6.&(3.12c)\cr}
$$
Set  
$$
\eqalignno{
\phi(k)&={1\over 2}\bigg[{k!(k+3)!\over (k+4)!}\bigg]^{{1\over 2}},&(3.13a)\cr
\sigma(k)&=\big[(k+1)!\big]^{1\over 2},&(3.13b)\cr
\tau(k)&={1\over 4}\bigg[{k!(k+2)!(k+3)!(k+5)! \over 
(k+4)!(k+7)!(k+10)!}\bigg]^{{1\over 2}}.&(3.13c)\cr}
$$
Then, 
$$
\eqalignno{
\bar R^{fff}_{I_1I_2I_3}&={\pi\over 2}\Big({2\over N}\Big)^{3\over 4}
\phi(k_1)\phi(k_2)\phi(k_3)
2^{-\alpha}
{\Gamma(\alpha+5)\over\Gamma(\alpha+3)}
{1\over\Gamma({1\over 2}\alpha+{5\over 2})}&\cr
&\times
{\Gamma(\alpha_1+2)\Gamma(\alpha_2+2)\Gamma(\alpha_3+2)
\over \Gamma(\alpha_1+1)\Gamma(\alpha_2+1)\Gamma(\alpha_3+1)}
{1\over \Gamma({1\over 2}\alpha_1+1)
\Gamma({1\over 2}\alpha_2+1)\Gamma({1\over 2}\alpha_3+1)}
&(3.14a)\cr
\bar R^{ffs}_{I_1I_2I_3}&={\pi\over 2}\Big({2\over N}\Big)^{3\over 4}
\phi(k_1)\phi(k_2)\sigma(k_3)
2^{-\alpha}{1\over\Gamma({1\over 2}\alpha+2)}&\cr
&\times
{\Gamma(\alpha_3+5)\over \Gamma(\alpha_3+1)}
{1\over\Gamma({1\over 2}(\alpha_1+1))\Gamma({1\over 2}(\alpha_2+1))
\Gamma({1\over 2}(\alpha_3+3))},
&(3.14b)\cr
\bar R^{fft}_{I_1I_2I_3}&={\pi\over 2}\Big({2\over N}\Big)^{3\over 4}
\phi(k_1)\phi(k_2)\tau(k_3)
2^{-\alpha}{\Gamma(\alpha+8)\over\Gamma(\alpha+4)}
{1\over\Gamma({1\over 2}\alpha+3)}&\cr
&\times{\Gamma(\alpha_1+5)\Gamma(\alpha_2+5)\over 
\Gamma(\alpha_1+1)\Gamma(\alpha_2+1)}
{1\over\Gamma({1\over 2}(\alpha_1+3))\Gamma({1\over 2}(\alpha_2+3))
\Gamma({1\over 2}(\alpha_3+1))},
&(3.14c)\cr
\bar R^{fss}_{I_1I_2I_3}&={\pi\over 2}\Big({2\over N}\Big)^{3\over 4}
\phi(k_1)\sigma(k_2)\sigma(k_3)
2^{-\alpha}{1\over\Gamma({1\over 2}\alpha+{3\over 2})}&\cr
&\times
{\Gamma(\alpha_2+2)\Gamma(\alpha_3+2)\over 
\Gamma(\alpha_2+1)\Gamma(\alpha_3+1)}
{1\over\Gamma({1\over 2}\alpha_1)\Gamma({1\over 2}\alpha_2+1)
\Gamma({1\over 2}\alpha_3+1)},
&(3.14d)\cr
\bar R^{fst}_{I_1I_2I_3}&={\pi\over 2}\Big({2\over N}\Big)^{3\over 4}
\phi(k_1)\sigma(k_2)\tau(k_3)
2^{-\alpha}{\Gamma(\alpha+5)\over\Gamma(\alpha+3)}
{1\over\Gamma({1\over 2}\alpha+{5\over 2})}&\cr
&\times
{\Gamma(\alpha_1+2)\over \Gamma(\alpha_1)}
{\Gamma(\alpha_2+8)\over \Gamma(\alpha_2+1)}
{\Gamma(\alpha_3)\over \Gamma(\alpha_3+1)}
{1\over\Gamma({1\over 2}\alpha_1+1)\Gamma({1\over 2}\alpha_2+2)
\Gamma({1\over 2}\alpha_3)},
&(3.14e)\cr
\bar R^{ftt}_{I_1I_2I_3}&={\pi\over 2}\Big({2\over N}\Big)^{3\over 4}
\phi(k_1)\tau(k_2)\tau(k_3)
2^{-\alpha}{\Gamma(\alpha+11)\over\Gamma(\alpha+3)}
{1\over\Gamma({1\over 2}\alpha+{7\over 2})}&\cr
&\times
{\Gamma(\alpha_1+8)\over \Gamma(\alpha_1)}
{\Gamma(\alpha_2+2)\over \Gamma(\alpha_2+1)}
{\Gamma(\alpha_3+2)\over \Gamma(\alpha_3+1)}
{1\over\Gamma({1\over 2}\alpha_1+2)\Gamma({1\over 2}\alpha_2+1)
\Gamma({1\over 2}\alpha_3+1)},
&(3.14f)\cr
\bar R^{sss}_{I_1I_2I_3}&={\pi\over 2}\Big({2\over N}\Big)^{3\over 4}
\sigma(k_1)\sigma(k_2)\sigma(k_3)
2^{-\alpha}{1\over\Gamma({1\over 2}\alpha+1)}&\cr
&\times{1\over 
\Gamma({1\over 2}(\alpha_1+1))\Gamma({1\over 2}(\alpha_2+1))
\Gamma({1\over 2}(\alpha_3+1))},
&(3.14g)\cr
\bar R^{sst}_{I_1I_2I_3}&={\pi\over 2}\Big({2\over N}\Big)^{3\over 4}
\sigma(k_1)\sigma(k_2)\tau(k_3)
2^{-\alpha}{1\over\Gamma({1\over 2}\alpha+2)}&\cr
&\times
{\Gamma(\alpha_1+5)\Gamma(\alpha_2+5)\over 
\Gamma(\alpha_1+1)\Gamma(\alpha_2+1)}
{1\over \Gamma({1\over 2}(\alpha_1+3))\Gamma({1\over 2}(\alpha_2+3))
\Gamma({1\over 2}(\alpha_3-1))},
&(3.14h)\cr
\bar R^{tts}_{I_1I_2I_3}&={\pi\over 2}\Big({2\over N}\Big)^{3\over 4}
\tau(k_1)\tau(k_2)\sigma(k_3)
2^{-\alpha}{\Gamma(\alpha+8)\over\Gamma(\alpha+4)}
{1\over\Gamma({1\over 2}\alpha+3)}&\cr
&\times
{\Gamma(\alpha_3+5)\Gamma(\alpha_3+11)\over 
\Gamma(\alpha_3+1)\Gamma(\alpha_3+3)}
{1\over \Gamma({1\over 2}(\alpha_1+1))
\Gamma({1\over 2}(\alpha_2+1))\Gamma({1\over 2}(\alpha_3+5))},
&(3.14i)\cr
\bar R^{ttt}_{I_1I_2I_3}&={\pi\over 2}\Big({2\over N}\Big)^{3\over 4}
\tau(k_1)\tau(k_2)\tau(k_3)
2^{-\alpha}
{\Gamma(\alpha+8)\Gamma(\alpha+14)\over\Gamma(\alpha+4)\Gamma(\alpha+6)}
{1\over\Gamma({1\over 2}\alpha+4)}&\cr
&\times
{\Gamma(\alpha_1+5)\Gamma(\alpha_2+5)\Gamma(\alpha_3+5)
\over \Gamma(\alpha_1+1)\Gamma(\alpha_2+1)\Gamma(\alpha_3+1)}
{1\over \Gamma({1\over 2}(\alpha_1+3))
\Gamma({1\over 2}(\alpha_2+3))\Gamma({1\over 2}(\alpha_3+3))}.
&(3.14j)\cr}
$$

\vskip.2cm

$\underline{{\rm AdS}_5\times{\rm S}_5}$

\vskip.2cm

In this case one has ${1\over 4\kappa^2} = {N^2 \over 8 \pi^5}$ and 
$\bar e=1$. One has
$$
\eqalignno{
\Delta^f_I&=k+4,&(3.15a)\cr
\Delta^s_I&=k,&(3.15b)\cr
\Delta^t_I&=k+8.&(3.15c)\cr}
$$
Set 
$$
\eqalignno{
\phi(k)&=\bigg[{k!\over (k+3)!}\bigg]^{{1\over 2}},&(3.16a)\cr
\sigma(k)&=k^{1\over 2},&(3.16b)\cr
\tau(k)&=\bigg[{k!(k+1)!(k+2)!\over 
(k+5)!(k+6)!(k+7)!}\bigg]^{{1\over 2}}.&(3.16c)\cr}
$$
Then, 
$$
\eqalignno{
\bar R^{fff}_{I_1I_2I_3}&={1\over N}\phi(k_1)\phi(k_2)\phi(k_3)
{\Gamma(\alpha+4)\over\Gamma(\alpha+2)}
{\Gamma(\alpha_1+2)\Gamma(\alpha_2+2)\Gamma(\alpha_3+2)
\over \Gamma(\alpha_1+1)\Gamma(\alpha_2+1)\Gamma(\alpha_3+1)}
&(3.17a)\cr
\bar R^{ffs}_{I_1I_2I_3}&={1\over N}
\phi(k_1)\phi(k_2)\sigma(k_3)
{\Gamma(\alpha_3+3)\Gamma(\alpha_3+4)\over
\Gamma(\alpha_3+1)\Gamma(\alpha_3+2)},
&(3.17b)\cr
\bar R^{fft}_{I_1I_2I_3}&={1\over N}
\phi(k_1)\phi(k_2)\tau(k_3)
{\Gamma(\alpha+5)\Gamma(\alpha+6)\over
\Gamma(\alpha+3)\Gamma(\alpha+4)}&\cr
&\times
{\Gamma(\alpha_1+3)\Gamma(\alpha_1+4)\over
\Gamma(\alpha_1+1)\Gamma(\alpha_1+2)}
{\Gamma(\alpha_2+3)\Gamma(\alpha_2+4)\over
\Gamma(\alpha_2+1)\Gamma(\alpha_2+2)},
&(3.17c)\cr
\bar R^{fss}_{I_1I_2I_3}&={1\over N}
\phi(k_1)\sigma(k_2)\sigma(k_3)
{\Gamma(\alpha_2+2)\over\Gamma(\alpha_2+1)}
{\Gamma(\alpha_3+2)\over\Gamma(\alpha_3+1)},
&(3.17d)\cr
\bar R^{fst}_{I_1I_2I_3}&={1\over N}
\phi(k_1)\sigma(k_2)\tau(k_3)
{\Gamma(\alpha+4)\over\Gamma(\alpha+2)}
&\cr
&\times
{\Gamma(\alpha_1+2)\over\Gamma(\alpha_1)}
{\Gamma(\alpha_2+5)\Gamma(\alpha_2+6)\over
\Gamma(\alpha_2+1)\Gamma(\alpha_2+3)}
{\Gamma(\alpha_3)\over\Gamma(\alpha_3+1)},
&(3.17e)\cr
\bar R^{ftt}_{I_1I_2I_3}&={1\over N}
\phi(k_1)\tau(k_2)\tau(k_3)
{\Gamma(\alpha+7)\Gamma(\alpha+8)\over
\Gamma(\alpha+2)\Gamma(\alpha+5)}&\cr
&\times
{\Gamma(\alpha_1+5)\Gamma(\alpha_1+6)\over
\Gamma(\alpha_1)\Gamma(\alpha_1+3)}
{\Gamma(\alpha_2+2)\over\Gamma(\alpha_2+1)}
{\Gamma(\alpha_3+2)\over\Gamma(\alpha_3+1)},
&(3.17f)\cr
\bar R^{sss}_{I_1I_2I_3}&={1\over N}
\sigma(k_1)\sigma(k_2)\sigma(k_3),
&(3.17g)\cr
\bar R^{sst}_{I_1I_2I_3}&={1\over N}
\sigma(k_1)\sigma(k_2)\tau(k_3)
{\Gamma(\alpha_1+3)\Gamma(\alpha_1+4) 
\Gamma(\alpha_2+3)\Gamma(\alpha_2+4)\over
\Gamma(\alpha_1+1)\Gamma(\alpha_1+2)
\Gamma(\alpha_2+1)\Gamma(\alpha_2+2)},
&(3.17h)\cr
\bar R^{tts}_{I_1I_2I_3}&={1\over N}
\tau(k_1)\tau(k_2)\sigma(k_3)
{\Gamma(\alpha+5)\Gamma(\alpha+6)\over\Gamma(\alpha+3)\Gamma(\alpha+4)}
{\Gamma(\alpha_3+7)\Gamma(\alpha_3+8)\over 
\Gamma(\alpha_3+1)\Gamma(\alpha_3+2)},
&(3.17i)\cr
\bar R^{ttt}_{I_1I_2I_3}&={1\over N}
\tau(k_1)\tau(k_2)\tau(k_3)
{\Gamma(\alpha+9)\Gamma(\alpha+10)\over\Gamma(\alpha+3)\Gamma(\alpha+4)}&\cr
&\times
{\Gamma(\alpha_1+3)\Gamma(\alpha_1+4)\Gamma(\alpha_2+3)\Gamma(\alpha_2+4)
\Gamma(\alpha_3+3)\Gamma(\alpha_3+4)
\over \Gamma(\alpha_1+1)\Gamma(\alpha_1+2)\Gamma(\alpha_2+1)\Gamma(\alpha_2+2)
\Gamma(\alpha_3+1)\Gamma(\alpha_3+2)}.
&(3.17j)\cr}
$$

It is useful to remember that the value of the integer $k$ of the operators
${\cal O}_I^s$, ${\cal O}_I^f$, ${\cal O}_I^t$ belonging to the same multiplet
are related as follows
$$
k^s=k,\quad\quad k^f=k-2,\quad\quad k^t=k-4.
\eqno(3.18)
$$

Some of these results have been obtained by other groups as well.
Namely, in the SCFT$_4$ case, 
the 3-point function of the chiral primary operators 
eq. (3.17$g$) was produced in  the seminal paper [13], 
while eq. (3.17$h$) was contained in [17].
As for SCFT$_6$, eq. (3.11$g$) was calculated in [15],
though that result differs for a factor from ours [23].
We have computed in previous works the $s$--$t$ correlations functions [23,24],
extended here above to the full universal sector.

\par\vskip.6cm
\item{\bf 4.} {\bf Conclusions}
\vskip.4cm

We have computed in a systematic way  the large $N$ limit of 
the 3-point functions of the scalars belonging to the universal sector
which is common to all of the maximal supersymmetric CFT in $d=3,4,6$.
The infrared dynamics of the M2, D3 and M5 branes belongs to 
the universality classes identified by these
SCFT$_{3,4,6}$, and in fact we used the AdS/CFT correspondence
to compute some scalar correlation functions in the latter.
That constitutes the main result of this paper.

At large $N$, all the operators group into short (also called chiral)
supersymmetric multiplets [31--35].
In particular, the operators ${\cal O}^s$ are the chiral primaries which
generate all of the other by application of the superconformal algebra
[36,37].
In $d=4$ it was shown in [38] by using a superspace approach 
that the 3-point function of all chiral operators are fixed by a unique
superspace conformal invariant up to a single coefficient, which can be 
easily read off from $\langle{\cal O}^s {\cal O}^s {\cal O}^s\rangle$.
Thus, it should be possible to obtain the correlation functions of 
the descendants, including the ones listed in 
section 3, by such superspace techniques,
though this may be laborious.
A similar strategy has not been worked out for the SCFT$_{3,6}$,
to our knowledge,
and it is not known how many independent numbers describe
the correlation functions of the chiral multiplets.
Investigation of the stress tensor 3-point functions 
(belonging to the family identified by the chiral primary with $k=2$)
show only a single overall coefficient for all of the SCFT$_{3,6}$
[39], and this may suggest that also in these new cases
 supersymmetry fixes the correlations function of chiral multiplets 
up to a constant.
It would indeed be interesting to develop a superspace approach 
and reproduce the descendant correlation functions from the chiral ones.
More importantly, it would be nice to investigate their 
$1\over N$ corrections. On the other hand, the general model we used 
to obtain the scalar 3-point function can still be employed to compute 
the 4-point functions of the chiral primaries
in a systematic way for the SCFT$_{3,4,6}$. 
In the light of some results for the super Yang Mills case [20], 
that might be a formidable task but still within the reach of 
future research.

\par\vskip.6cm
\item{\bf A1.} {\bf Spherical Harmonics}
\vskip.4cm

\par 
We describe the $n$--dimensional sphere ${\rm S}_n$ of radius 
$\rho$ as ${\rm S}_n=\{x\in\Bbb R^{n+1}|x^2=\rho^2\}$.

Since ${\rm S}_n\subset\Bbb R^{n+1}$, we can represent tensor fields 
on ${\rm S}_n$ by means of tensor fields on $\Bbb R^{n+1}$ of the same type 
satisfying certain conditions. Since further ${\rm S}_n$ is naturally equipped 
with the metric induced by the Euclidean metric of $\Bbb R^{n+1}$, 
we need not distinguish covariant and contravariant tensor indices. 
Specifically, a rank $s$ tensor field $T_{\alpha_1\ldots\alpha_s}$
on ${\rm S}_n$ may be viewed as a rank $s$ tensor field $T_{i_1\ldots i_s}$
defined on a neighborhood of ${\rm S}_n\subset\Bbb R^{n+1}$ 
with no components normal to ${\rm S}_n$: 
$$
x_i T_{i_1\ldots i_{a-1}ii_{a+1}\ldots i_s}(x)=0,\quad1\leq a\leq s.
\eqno(A1.1)
$$
In particular, the induced metric $g_{\alpha\beta}$ of ${\rm S}_n$ is 
represented by
$$
g_{ij}(x)=\delta_{ij}-{x_ix_j\over r^2},
\eqno(A1.2)
$$
where $r=(x_ix_i)^{1\over 2}$. Therefore, the contraction of two rank 
$s$ tensor fields $T_{\alpha_1\ldots\alpha_s}$, $U_{\alpha_1\ldots\alpha_s}$
is given by 
$$
T^{\alpha_1\ldots\alpha_s}U_{\alpha_1\ldots\alpha_s}
=T_{i_1\ldots i_s}U_{i_1\ldots i_s}.
\eqno(A1.3)
$$

For a rank $s$ tensor field $T_{\alpha_1\ldots\alpha_s}$,
the covariant derivative
$\nabla_\alpha T_{\alpha_1\ldots\alpha_s}$ is represented by 
$$
\nabla_i T_{i_1\ldots i_s}(x)
=\Big(\partial_i-{x_i\over r}\partial_r\Big) T_{i_1\ldots i_s}(x)
+\sum_{a=1}^s{x_{i_a}\over r^2}T_{i_1\ldots i_{a-1}ii_{a+1}\ldots i_s}(x),
\eqno(A1.4)
$$
where $\partial_r=\partial/\partial r$.

For any function $F$ on ${\rm S}_n$, the integral of $F$ on ${\rm S}_n$
is given by
$$
\int_{{\rm S}_n} d^n{\rm vol} F
=\int_{{\rm S}_n} r^n d\Omega_n(x)F(x)\bigg |_{r=\rho},
\eqno(A1.5)
$$
where $d\Omega_n(x)$ is the standard $n$ dimensional volume
on ${\rm S}_n$ defined as usual by $d^{n+1}x=r^n dr d\Omega_n(x)$.

Using the above integration formula, one can provide the space of rank $s$ 
tensor fields with a Hilbert space structure in obvious fashion. 

Consider the rank $s$ tensor fields 
$$
Y^{(s)}_{Ii_1\ldots i_s}(x)={\cal Y}^{(s)}_{Ii_1\ldots i_s;j_1\ldots j_k}
r^{-k} x_{j_1}\ldots x_{j_k},
\eqno(A1.6)
$$
where the ${\cal Y}^{(s)}_{Ii_1\ldots i_s;j_1\ldots j_k}$ are constants
satisfying the following conditions:

\item{$i)$} ${\cal Y}^{(s)}_{Ii_1\ldots i_s;j_1\ldots j_k}$ is symmetric 
and traceless in $i_1,\ldots, i_s$;

\item{$ii)$} ${\cal Y}^{(s)}_{Ii_1\ldots i_s;j_1\ldots j_k}$ is symmetric 
and traceless in $j_1,\ldots, j_k$;

\item{$iii)$} the following relation holds 
$$
{\cal Y}^{(s)}_{Ii_1\ldots i_{s-1}\{i_s;j_1\ldots j_k\}}=0,
\eqno(A1.7)
$$
where $\{\ldots\}$ denotes total symmetrization.

\noindent
The index $I$ labels the different choices of the constants
${\cal Y}^{(s)}_{Ii_1\ldots i_s;j_1\ldots j_k}$. 
Note that $Y^{(s)}_{Ii_1\ldots i_s}$ is characterized by the non negative 
integer $k$, which may be thought of as a function of the label $I$.
It can be shown that, if $Y^{(s)}_{Ii_1\ldots i_s}\not=0$, then 
necessarily $k\geq s$.

Then, the $Y^{(s)}_{Ii_1\ldots i_s}$ are symmetric traceless rank $s$
tensor fields on ${\rm S}_n$. Further, they are divergenceless (transversal):
$$
\nabla_iY^{(s)}_{Iii_2\ldots i_s}=0,
\eqno(A1.8)
$$
Finally, the $Y^{(s)}_{Ii_1\ldots i_s}$ are eigenfields of the Laplacian:
$$
-\nabla_i\nabla_iY^{(s)}_{Ii_1\ldots i_s}
={1\over r^2}\big[(k(k+n-1)-s]Y^{(s)}_{Ii_1\ldots i_s}.
\eqno(A1.9)
$$

If, for fixed $s$, we choose a maximal set of constants 
${\cal Y}^{(s)}_{Ii_1\ldots i_s;j_1\ldots j_k}$ such that
$$
{\cal Y}^{(s)}_{Ii_1\ldots i_s;j_1\ldots j_k}
{\cal Y}^{(s)}_{Ji_1\ldots i_s;j_1\ldots j_k}=\delta_{IJ}
\eqno(A1.10)
$$
for all $k$, then, the tensor fields $\{Y^{(s)}_{Ii_1\ldots i_s}|I\}$ 
provide an orthogonal basis of the space of divergenceless symmetric 
traceless rank $s$ tensor fields on ${\rm S}_n$.

In concrete applications, one has to compute the integrals of scalars formed 
by contraction of several tensor spherical harmonics 
$Y^{(s_\ell)}$ and their covariant derivatives.
This can be done by a systematic use of the formula
$$
\eqalignno{
\int_{S^n} r^n d\Omega_n(x) r^{-2m} x^{i_1} \cdots x^{i_{2m}}\bigg |_{r=\rho} 
&=\rho^n\omega_n {(n-1)!! \over (2m+n-1)!!}&\cr 
&\times (\hbox{all possible Wick contractions}),
&(A1.11)\cr}
$$
where ``all possible Wick contractions'' are given by the sum of
$(2m -1)!!$ terms obtained by using 
$\langle r^{-2} x^i x^j\rangle=\delta^{ij}$ 
as elementary Wick contraction and 
$\omega_n$ is the volume of the unit sphere
$$
\omega_n={2\pi^{n+1\over 2}\over\Gamma({n+1\over2})}.
\eqno(A1.12)
$$
The results are always expressible in terms of suitable contractions of the 
coefficients ${\cal Y}^{(s_\ell)}_{Ii_1\ldots i_{s_\ell};j_1
\ldots j_{k_\ell}}$ generically denoted as $\langle \prod_\ell 
{\cal Y}^{(s_\ell)}_{I_\ell}\rangle$ and certain numerical functions
of the $k_\ell$.

For two and three tensor spherical harmonics, the only such functions are
$$
\eqalignno{
z_I&= \omega_n {(n-1)!! k!\over{(2k+n-1)!!}},
&(A1.13)\cr
a_{I_1I_2I_3}&=\omega_n  {(n-1)!!\over{(2\alpha+n-1)!!}}
{k_1!k_2!k_3! \over \alpha_1! \alpha_2! \alpha_3!}.
&(A1.14)\cr}
$$
where
$$
\alpha_1=\hbox{$1\over 2$}(k_2+k_3-k_1), \quad
\alpha_2=\hbox{$1\over 2$}(k_3+k_1-k_2), \quad
\alpha_3=\hbox{$1\over 2$}(k_1+k_2-k_3),  
\eqno(A1.15a)
$$
$$
\alpha=\hbox{$1\over 2$}(k_1+k_2+k_3).
\eqno(A1.15b)
$$

In this paper, $\rho=\bar e^{-1}$. Further, we consider only $s=0,2$.
We set ${\cal C}_{I;j_1\ldots j_k}={\cal Y}^{(0)}_{I;j_1\ldots j_k}$
and ${\cal T}_{Ii_1i_2;j_1\ldots j_k}={\cal Y}^{(2)}_{Ii_1i_2;j_1\ldots j_k}$.

The main contractions are the following.
$\langle{\cal C}_{I_1}{\cal C}_{I_2}{\cal C}_{I_3}\rangle$ 
denotes the unique $SO(n+1)$ scalar contraction
of three tensors  ${\cal C}_{Ii_1...i_k}$ and in 
a similar fashion we define
$$
\eqalignno{
\langle{\cal T}_{I_1}{\cal C}_{I_2}{\cal C}_{I_3}\rangle &=
\langle{\cal T}_{I_1ab}{\cal C}_{I_2;a}{\cal C}_{I_3;b}\rangle 
&(A1.16a)\cr
\langle{\cal T}_{I_1}{\cal T}_{I_2}{\cal C}_{I_3}\rangle &=
\langle{\cal T}_{I_1ab}{\cal T}_{I_2ab}{\cal C}_{I_3}\rangle 
&(A1.16b)\cr
\langle{\cal T}_{I_1}{\cal T}_{I_2}{\cal T}_{I_3}\rangle
& =4\langle{\cal T}_{I_1ab}{\cal T}_{I_2bc}{\cal T}_{I_3ca}\rangle 
\cr &
+ \sum_{ {\rm c.p.} 123} \alpha_1\Big( 
2\langle{\cal T}_{I_1ab}{\cal T}_{I_2ac;d}{\cal T}_{I_3bd;c}\rangle
+\langle{\cal T}_{I_1ab}{\cal T}_{I_2cd;a}{\cal T}_{I_3cd;b}\rangle\Big)
&(A1.16c)\cr}
$$
where on the right hand side one takes the unique 
contraction of the hidden indices, and 
where the notation ${\rm c.p.}123$ stands for cyclic 
permutations of $123$.

\par\vskip.6cm
\item{\bf A2.} {\bf Auxiliary functions }
\vskip.4cm
\par

Here is the complete list of the auxiliary functions 
$F$ appearing in $(2.4a)-(2.4j)$.
$$
\eqalignno{
F^{fss} 
&=  (D - 2)\alpha_2\alpha_3  &(A2.1a)\cr
F^{ftt} 
&=(D - 2)\alpha_2\alpha_3  &(A2.1b)\cr
F^{fst} 
&= (D - 2)\alpha_1 (\alpha + 1 + p)  &(A2.1c)\cr
F^{sss}
&=(1+p)(D-3-p)
(\alpha_1{}^2 \alpha_2 + \alpha_2{}^2 \alpha_1 + \alpha_2{}^2 \alpha_3 
+ \alpha_3{}^2 \alpha_2 + \alpha_3{}^2 \alpha_1 + \alpha_1{}^2 \alpha_3)
\vphantom{2\over 3}&\cr
&\hphantom{~~\times\bigg\{}
+ (3D-8 + (2D-8)p - 2p^2)\alpha_1 \alpha_2 \alpha_3 
\vphantom{2\over 3}&\cr
&\hphantom{~~\times\bigg\{}
+ (1+p)(-\hbox{$1\over 2$}D+2+p)(\alpha_1\alpha_2 +\alpha_2 \alpha_3 
+ \alpha_3 \alpha_1) &(A2.1d)\cr
F^{sst}
&=(1+p)(D-3-p)  
(\alpha_1{}^2 \alpha_3 + \alpha_3{}^2 \alpha_1 
+ \alpha_2{}^2 \alpha_3 + \alpha_3{}^2 \alpha_2)
\vphantom{2\over 3}&\cr
&\hphantom{~~\times\bigg\{}  
+ (-1 + (D-4) p - p^2) (\alpha_1{}^2 \alpha_2 +\alpha_2{}^2 \alpha_1) 
\vphantom{2\over 3}&\cr
&\hphantom{~~\times\bigg\{}  
+ (D-4 + (2D-8) p - 2 p^2) \alpha_1 \alpha_2 \alpha_3 
\vphantom{2\over 3}&\cr
&\hphantom{~~\times\bigg\{}
+ (1+p) p (D-3-p)(\alpha_1 \alpha_3 + \alpha_2 \alpha_3) 
\vphantom{2\over 3}&\cr
&\hphantom{~~\times\bigg\{}
+ (1+p) (-D+2 + (D-3) p - p^2) \alpha_1 \alpha_2 
\vphantom{2\over 3}&\cr
&\hphantom{~~\times\bigg\{} 
+(1+p) (D-3-p) (-\alpha_1{}^2 - \alpha_2{}^2 + (1+p)\alpha_3{}^2)
\vphantom{2\over 3}&\cr
&\hphantom{~~\times\bigg\{}
+ (1+p)^2 (-D+3+p) (\alpha_1 + \alpha_2 +\alpha_3) 
&(A2.1e)\cr
F^{tts}
&=(1+p)(D-3-p) (\alpha_1{}^2 \alpha_3 + \alpha_3{}^2 \alpha_1 
+ \alpha_2{}^2 \alpha_3 + \alpha_3{}^2 \alpha_2)
\vphantom{2\over 3}&\cr
&\hphantom{~~\times\bigg\{}  
+  (-1 + (D-4) p - p^2) (\alpha_1{}^2 \alpha_2 +\alpha_2{}^2 \alpha_1) 
\vphantom{2\over 3}&\cr
&\hphantom{~~\times\bigg\{}  
+  (D-4 + (2D-8) p - 2 p^2) \alpha_1 \alpha_2 \alpha_3 
\vphantom{2\over 3}&\cr
&\hphantom{~~\times\bigg\{}
+ (1+p) (\hbox{$5\over 2$}D-8 + (2D- 9) p - 2 p^2)
(\alpha_1 \alpha_3 + \alpha_2 \alpha_3) 
\vphantom{2\over 3}&\cr
&\hphantom{~~\times\bigg\{}
+ (1+p) (\hbox{$3\over 2$}D-6 + (2D-9) p - 2 p^2) \alpha_1 \alpha_2 
\vphantom{2\over 3}&\cr
&\hphantom{~~\times\bigg\{} 
+ (1+p) (\hbox{$3\over 2$}D-5 + (D-5) p -  p^2) (\alpha_1{}^2 + \alpha_2{}^2)
\vphantom{2\over 3}&\cr
&\hphantom{~~\times\bigg\{}
+ (1+p)^2 (\hbox{$3\over 2$}D -5 + (D-5) p -  p^2) (\alpha_1 + \alpha_2)
&(A2.1f)\cr
F^{ttt} &=(1+p)(D-3-p)
(\alpha_1{}^2 \alpha_2 + \alpha_2{}^2 \alpha_1 + \alpha_2{}^2 \alpha_3 
+ \alpha_3{}^2 \alpha_2 + \alpha_3{}^2 \alpha_1 + \alpha_1{}^2 \alpha_3)
\vphantom{2\over 3}&\cr
&\hphantom{~~\times\bigg\{}
+ (3D-8 + (2D-8)p - 2p^2)\alpha_1 \alpha_2 \alpha_3 
\vphantom{2\over 3}&\cr
&\hphantom{~~\times\bigg\{}
+ (1+p)(4D-13 + (3D-14)p - 3p^2)(\alpha_1\alpha_2 +\alpha_2 \alpha_3 
+ \alpha_3 \alpha_1)
\vphantom{2\over 3}&\cr
&\hphantom{~~\times\bigg\{}
+ (1+p)^2 (D -\hbox{$7\over 2$} -p)(\alpha_1{}^2 +\alpha_2{}^2 +\alpha_3{}^2)
\vphantom{2\over 3}&\cr
&\hphantom{~~\times\bigg\{}
+ (1+p)^2 (3D -11  +(2D-\hbox{$21\over 2$})p -2 p^2)
(\alpha_1+\alpha_2+\alpha_3)
\vphantom{2\over 3}&\cr
&\hphantom{~~\times\bigg\{}
+(2 +p)(1+p)^3 (D-4-p) 
&(A2.1g)\cr}
$$

\vfill\eject

\centerline{\bf REFERENCES}

\item{[1]} N.~Seiberg,
``Notes on theories with 16 supercharges'',
Nucl.\ Phys.\ Proc.\ Suppl.\  {\bf 67} (1998) 158,   
{\tt hep-th/9705117}.

\item{[2]} E.~Witten,
``Bound states of strings and p-branes'',
Nucl.\ Phys.\  {\bf B460} (1996) 335,
{\tt hep-th/9510135}.

\item{[3]} G.~W.~Gibbons and P.~K.~Townsend,
``Vacuum interpolation in supergravity via super p-branes'',
Phys.\ Rev.\ Lett.\  {\bf 71} (1993) 3754,
{\tt hep-th/9307049}.

\item{[4]} D.~M.~Kaplan and J.~Michelson,
``Zero modes for the D=11 membrane and five-brane'',
Phys.\ Rev.\  {\bf D53} (1996) 3474,
{\tt hep-th/9510053}.

\item{[5]} E.~Bergshoeff, E.~Sezgin and P.~K.~Townsend,
``Supermembranes and eleven dimensional supergravity'',
Phys.\ Lett.\  {\bf B189} (1987) 75.

\item{[6]} E.~Witten,
``Some comments on string dynamics'',
{\tt hep-th/9507121}.

\item{[7]} A.~Strominger,
``Open p-branes'',
Phys.\ Lett.\  {\bf B383} (1996) 44,
{\tt hep-th/9512059}.

\item{[8]} E.~Witten,
``Five-branes and M-theory on an orbifold'',
Nucl.\ Phys.\  {\bf B463} (1996) 383,
{\tt hep-th/9512219}.

\item{[9]} J.~Maldacena,
``The large $N$ limit of superconformal field theories and supergravity'',
Adv.\ Theor.\ Math.\ Phys.\  {\bf 2} (1998) 231,
{\tt hep-th/9711200}.

\item{[10]} S.~S.~Gubser, I.~R.~Klebanov and A.~M.~Polyakov,
``Gauge theory correlators from non-critical string theory'',
Phys.\ Lett.\  {\bf B428} (1998) 105,
{\tt hep-th/9802109}.

\item{[11]} E.~Witten,
``Anti-de Sitter space and holography'',
Adv.\ Theor.\ Math.\ Phys.\  {\bf 2} (1998) 253,
{\tt hep-th/9802150}.
 
\item{[12]} O.~Aharony, S.~S.~Gubser, J.~Maldacena, H.~Ooguri and Y.~Oz,
``Large $N$ field theories, string theory and gravity'',
Phys.\ Rept.\  {\bf 323} (2000) 183,
{\tt hep-th/9905111}.

\item{[13]} S.~Lee, S.~Minwalla, M.~Rangamani and N.~Seiberg,
``Three-point functions of chiral operators in $D=4$, 
${\cal N}=4$ SYM at Large $N$'', 
Adv.\ Theor.\ Math.\ Phys.\  {\bf 2} (1998) 697,
{\tt hep-th/9806074}.

\item{[14]} G.~Arutyunov and S.~Frolov,
``Three-point Green function of the stress-energy tensor in the 
AdS/CFT  correspondence'',
Phys.\ Rev.\  {\bf D60} (1999) 026004,
{\tt hep-th/9901121}.

\item{[15]} R.~Corrado, B.~Florea and R.~McNees,
``Correlation functions of operators and Wilson surfaces in the  
$d=6$, $(0,2)$ theory in the large $N$ limit'',
Phys.\ Rev.\  {\bf D60} (1999) 085011,
{\tt hep-th/9902153}.

\item{[16]} E.~D'Hoker, D.~Z.~Freedman, S.~D.~Mathur, 
A.~Matusis and L.~Rastelli,
``Graviton exchange and complete 4-point functions in the AdS/CFT  
correspondence'',
Nucl.\ Phys.\  {\bf B562} (1999) 353,
{\tt hep-th/9903196}.

\item{[17]} G.~Arutyunov and S.~Frolov,
``Some cubic couplings in type IIB supergravity on $AdS_5\times S^5$ 
and  three-point functions in SYM$_4$ at large $N$'',
Phys.\ Rev.\  {\bf D61} (2000) 064009,
{\tt hep-th/9907085}.

\item{[18]} S.~Lee,
``AdS(5)/CFT(4) four-point functions of chiral primary operators: 
cubic  vertices'',
Nucl.\ Phys.\  {\bf B563} (1999) 349,
{\tt hep-th/9907108}.

\item{[19]} E.~D'Hoker, D.~Z.~Freedman, S.~D.~Mathur, A.~Matusis 
and L.~Rastelli,
``Extremal correlators in the AdS/CFT correspondence'',
{\tt hep-th/9908160}.

\item{[20]} G.~Arutyunov and S.~Frolov,
``Four-point functions of lowest weight CPOs in ${\cal N}=4$
 SYM$_4$ in  supergravity approximation'',
{\tt hep-th/0002170}.

\item{[21]} R.~Manvelian and A.~C.~Petkou,
``A note on R-currents and trace anomalies in the $(2,0)$ tensor multiplet  
in $d=6$ AdS/CFT correspondence'',
{\tt hep-th/0003017}.

\item{[22]} F.~Bastianelli and R.~Zucchini,
``Bosonic quadratic actions for 11D supergravity on
${\rm AdS}_{7/4} \times {\rm S}_{4/7}$'',
Class.\ Quant.\ Grav.\  {\bf 16} (1999) 3673,
{\tt hep-th/9903161}.

\item{[23]} F.~Bastianelli and R.~Zucchini,
``Three point functions of chiral primary operators in 
$d=3$, ${\cal N} = 8$ and  $d=6$, ${\cal N}=(2,0)$ SCFT at large $N$'',
Phys.\ Lett.\  {\bf B467} (1999) 61,
{\tt hep-th/9907047}.

\item{[24]} F.~Bastianelli and R.~Zucchini,
``Three point functions for a class of chiral operators in maximally  
supersymmetric CFT at large $N$'',
hep-th/9909179.

\item{[25]} K.~Intriligator,
``Bonus symmetries of ${\cal N} = 4$ super-Yang-Mills correlation functions 
via  AdS duality'',
Nucl.\ Phys.\  {\bf B551} (1999) 575,
{\tt hep-th/9811047}.

\item{[26]} K.~Intriligator,
``Maximally supersymmetric RG flows and AdS duality,''
{\tt hep-th/ 9909082}.

\item{[27]} P. Freund and M. Rubin, 
``Dynamics of dimensional reduction'',
Phys. Lett. {\bf B97} (1980), 233.

\item{[28]} P. van Nieuwenhuizen,
``The complete mass spectrum of $d=11$ supergravity
compactified on $S^4$ and a general mass formula 
for arbitrary cosets $M_4$'', Class. Quantum Grav. {\bf 2} (1985), 1.

\item{[29]} W.~Muck and K.~S.~Viswanathan,
``Conformal field theory correlators from classical scalar field theory  
on AdS(d+1)'',
Phys.\ Rev.\  {\bf D58} (1998) 041901,
{\tt hep-th/9804035}.

\item{[30]} D.~Z.~Freedman, S.~D.~Mathur, A.~Matusis and L.~Rastelli,
``Correlation functions in the CFT($d$)/AdS($d+1$) correspondence'',
Nucl.\ Phys.\  {\bf B546} (1999) 96,
{\tt hep-th/ 9804058}.

\item{[31]} M.~Gunaydin, P.~van Nieuwenhuizen and N.~P.~Warner,
``General construction of the unitary representations 
of anti-De Sitter superalgebras and the spectrum of 
the $S^4$ compactification of eleven-dimensional supergravity'',
Nucl.\ Phys.\  {\bf B255} (1985) 63.

\item{[32]} M.~Gunaydin and N.~Marcus,
``The spectrum of the $S^5$ compactification of the chiral $N=2$, 
$D = 10$ supergravity and the unitary supermultiplets of $U(2, 2/4)$'',
Class.\ Quant.\ Grav.\  {\bf 2} (1985) L11.

\item{[33]} M.~Gunaydin, D.~Minic and M.~Zagermann,
``Novel supermultiplets of $SU(2,2|4)$ and the AdS(5)/CFT(4) duality'',
Nucl.\ Phys.\  {\bf B544} (1999) 737,
{\tt hep-th/9810226};
``4D doubleton conformal theories, CPT and II B string on AdS(5) x S(5)'',
Nucl.\ Phys.\  {\bf B534} (1998) 96
{\tt hep-th/9806042}.

\item{[34]} S.~Ferrara and E.~Sokatchev,
``Representations of (1,0) and (2,0) superconformal algebras in six  
dimensions: Massless and short superfields'',
{\tt hep-th/0001178};
``Short representations of SU(2,2/N) and harmonic superspace analyticity'',
{\tt hep-th/9912168}.

\item{[35]} L.~Andrianopoli, S.~Ferrara, E.~Sokatchev and B.~Zupnik,
``Shortening of primary operators in N-extended SCFT(4) and  
harmonic-superspace analyticity'',
{\tt hep-th/ 9912007}.

\item{[36]} W.~Nahm,
``Supersymmetries and their representations'',
Nucl.\ Phys.\  {\bf B135} (1978) 149.

\item{[37]} V.~K.~Dobrev and V.~B.~Petkova,
``All positive energy unitary irreducible representations 
of extended conformal supersymmetry'',
Phys.\ Lett.\  {\bf B162} (1985) 127;
``Group theoretical approach to extended conformal supersymmetry:
 function space realizations and invariant differential operators'',
Fortsch.\ Phys.\  {\bf 35} (1987) 537.

\item{[38]} P.~S.~Howe, E.~Sokatchev and P.~C.~West,
``Three point functions in $N=4$ Yang-Mills'', 
Phys.\ Lett.\  {\bf B444} (1998) 341,
{\tt hep-th/9808162}.

\item{[39]} F.~Bastianelli, S.~Frolov and A.~A.~Tseytlin,
``Three-point correlators of stress tensors in maximally-supersymmetric  
conformal theories in d = 3 and d = 6'', 
{\tt hep-th/9911135};
``Conformal anomaly of (2,0) tensor multiplet in six dimensions and 
 AdS/CFT correspondence'', JHEP {\bf 02} (2000) 013,
{\tt hep-th/0001041}.

\bye